\documentstyle[prd,aps,preprint,tighten,epsfig]{revtex}

\begin{document}

\draft

\title{Why is the $3\times 3$ neutrino mixing matrix almost unitary \\
in realistic seesaw models?}
\author{{\bf Zhi-zhong Xing}
\thanks{E-mail: xingzz@mail.ihep.ac.cn} ~ and ~ {\bf Shun Zhou}
\thanks{E-mail: zhoush@mail.ihep.ac.cn}}
\address{CCAST (World Laboratory), P.O. Box 8730, Beijing 100080,
China \\
and Institute of High Energy Physics, Chinese Academy of Sciences, \\
P.O. Box 918, Beijing 100049, China}
\maketitle

\begin{abstract}
A simple extension of the standard model is to introduce $n$ heavy
right-handed Majorana neutrinos and preserve its $\rm SU(2)^{}_L
\times U(1)^{}_Y$ gauge symmetry. Diagonalizing the $(3+n) \times
(3+n)$ neutrino mass matrix, we obtain an exact analytical
expression for the effective mass matrix of $\nu^{}_e$,
$\nu^{}_\mu$ and $\nu^{}_\tau$. It turns out that the $3\times 3$
neutrino mixing matrix $V$, which appears in the leptonic
charged-current weak interactions, must not be exactly unitary.
The unitarity violation of $V$ is negligibly tiny, however, if the
canonical seesaw mechanism works to reproduce the correct mass
scale of light Majorana neutrinos. A similar conclusion can be
drawn in the realistic Type-II seesaw models.
\end{abstract}


\newpage

\framebox{\large\bf 1} ~ Recent solar \cite{SNO}, atmospheric
\cite{SK}, reactor \cite{KM} and accelerator \cite{K2K} neutrino
oscillation experiments have provided us with very robust evidence
that neutrinos are massive and lepton flavors are mixed. This
great breakthrough opens a new window to physics beyond the
standard model (SM). Indeed, the fact that the masses of neutrinos
are considerably smaller than those of charged leptons and quarks
remains a big puzzle in particle physics. Although a lot of
theoretical models about the origin of neutrino masses have been
proposed at either low or high energy scales \cite{Review}, none
of them has proved to be very successful and conceivable.

Within the SM, neutrinos are massless particles and lepton flavor
mixing does not exist. The flavor eigenstates of three charged
leptons ($e$, $\mu$, $\tau$) and three neutrinos ($\nu^{}_e$,
$\nu^{}_\mu$, $\nu^{}_\tau$), which appear in the leptonic
charged-current weak interactions
\begin{equation}
-{\cal L}^{}_{\rm cc} \; =\; \frac{g}{\sqrt{2}} ~ \overline{(e,
~\mu, ~\tau)^{~}_{\rm L}} ~ \gamma^\mu \left (\matrix{ \nu^{}_e
\cr \nu^{}_\mu \cr \nu^{}_\tau \cr} \right )_{\rm L} W^-_\mu ~ + ~
{\rm h.c.} \; ,
\end{equation}
can therefore be identified with their corresponding mass
eigenstates. Beyond the SM, neutrinos may gain tiny but
non-vanishing masses through certain new interactions at low or
high energy scales. In this case, there is the phenomenon of
lepton flavor mixing in analogy with that of quark flavor mixing.
Identifying the flavor eigenstates of charged leptons with their
mass eigenstates, we may express $\nu^{}_e$, $\nu^{}_\mu$ and
$\nu^{}_\tau$ in terms of their mass eigenstates $\nu^{}_1$,
$\nu^{}_2$ and $\nu^{}_3$ as follows:
\begin{equation}
\left (\matrix{ \nu^{}_e \cr \nu^{}_\mu \cr \nu^{}_\tau \cr}
\right ) \; =\; \left (\matrix{ V^{}_{e1} & V^{}_{e2} & V^{}_{e3}
\cr V^{}_{\mu 1} & V^{}_{\mu 2} & V^{}_{\mu 3} \cr V^{}_{\tau 1} &
V^{}_{\tau 2} & V^{}_{\tau 3} \cr} \right ) \left (\matrix{
\nu^{}_1 \cr \nu^{}_2 \cr \nu^{}_3 \cr} \right ) \; .
\end{equation}
The transformation matrix $V$ in Eq. (2) is just the $3\times 3$
lepton flavor mixing matrix, sometimes referred to as the
Maki-Nakagawa-Sakata (MNS) matrix \cite{MNS}. Unlike the
Cabibb-Kobayashi-Maskawa (CKM) quark mixing matrix \cite{CKM},
which is required to be unitary in the SM, the MNS matrix $V$
comes from new physics beyond the SM and its unitarity is {\it
not} necessarily guaranteed in a specific model. If neutrinos are
Majorana particles and $V$ is exactly unitary, one can parametrize
$V$ in terms of three mixing angles and three CP-violating phases
\cite{Xing04}. If the unitarity of $V$ were significantly
violated, more free parameters would in general be needed to
describe neutrino mixing. A stringent test of the unitarity of $V$
turns out to be one of the most important goals in the future
neutrino factories and super-beam facilities.

The main purpose of this short paper is to show why the $3\times
3$ MNS matrix $V$ is not exactly unitary in a variety of neutrino
models incorporated with the famous seesaw mechanism \cite{SS}. To
be explicit, we extend the SM by including $n$ heavy right-handed
Majorana neutrinos and keeping its $\rm SU(2)^{}_L \times
U(1)^{}_Y$ gauge symmetry invariant. After diagonalizing the
$(3+n) \times (3+n)$ neutrino mass matrix, we arrive at an exact
analytical expression for the effective mass matrix of $\nu^{}_e$,
$\nu^{}_\mu$ and $\nu^{}_\tau$. Then it becomes obvious that the
MNS matrix $V$, which appears in the leptonic charged-current weak
interactions, is not exactly unitary. We find that the unitarity
violation of $V$ is negligibly tiny, unless the canonical seesaw
mechanism fails to reproduce the correct mass scale of light
Majorana neutrinos. A similar conclusion can be drawn in the
realistic Type-II seesaw mechanism.

\vspace{0.5cm}

\framebox{\large\bf 2} ~ Let us make a simple extension of the SM
by introducing $n$ heavy right-handed Majorana neutrinos $N^{}_i$
(for $i=1, \cdots, n$) and keeping the Lagrangian of electroweak
interactions invariant under the $\rm SU(2)^{}_L \times U(1)^{}_Y$
gauge transformation. In this case, the Lagrangian relevant for
lepton masses can be written as
\begin{equation}
-{\cal L}^{}_{\rm lepton} \; =\; \overline{l^{}_{\rm L}} Y^{}_l
e^{}_{\rm R} H + \overline{l^{}_{\rm L}} Y^{}_\nu N^{}_{\rm R}
H^{\rm c} + \frac{1}{2} \overline{N^{\rm c}_{\rm R}} M^{}_{\rm R}
N^{}_{\rm R} + {\rm h.c.} \; ,
\end{equation}
where $l^{}_{\rm L}$ denotes the left-handed lepton doublets;
$e^{}_{\rm R}$ and $N^{}_{\rm R}$ stand respectively for the
right-handed charged-lepton and Majorana neutrino singlets; $H$ is
the Higgs-boson weak isodoublet (with $H^{\rm c} \equiv
i\sigma^{}_2 H^*$); $M^{}_{\rm R}$ is the heavy Majorana neutrino
mass matrix; $Y^{}_l$ and $Y^{}_\nu$ are the coupling matrices of
charged-lepton and neutrino Yukawa interactions. After spontaneous
gauge symmetry breaking, the neutral component of $H$ acquires the
vacuum expectation value $v \approx 174$ GeV. Then we arrive at
the charged-lepton mass matrix $M^{}_l = v Y^{}_l$ and the
Dirac-type neutrino mass matrix $M^{}_{\rm D} = v Y^{}_\nu$. The
overall lepton mass term turns out to be
\begin{equation}
-{\cal L}^\prime_{\rm lepton} \; =\; \overline{e^{}_{\rm L}}
M^{}_l e^{}_{\rm R} + \frac{1}{2} \overline{(\nu^{}_{\rm L},
~N^{\rm c}_{\rm R} )} \left ( \matrix{ {\bf 0}     & M^{}_{\rm D}
\cr M^T_{\rm D}    & M^{}_{\rm R} \cr} \right ) \left ( \matrix{
\nu^{\rm c}_{\rm L} \cr N^{}_{\rm R} \cr} \right ) + {\rm h.c.} \;
,
\end{equation}
where $e$, $\nu^{}_{\rm L}$ and $N^{}_{\rm R}$ represent the
column vectors of $(e, \mu, \tau)$, $(\nu^{}_e, \nu^{}_\mu,
\nu^{}_\tau)^{}_{\rm L}$ and $(N^{}_\alpha, N^{}_\beta, \cdots
)^{}_{\rm R}$ fields, respectively. In obtaining Eq. (4), we have
used the relation $\overline{\nu^{}_{\rm L}} M^{}_{\rm D}
N^{}_{\rm R} = \overline{N^{\rm c}_{\rm R}} M^T_{\rm D} \nu^{\rm
c}_{\rm L}$ as well as the properties of $\nu^{}_{\rm L}$ (or
$N^{}_{\rm R}$) and $\nu^{\rm c}_{\rm L}$ (or $N^{\rm c}_{\rm R}$)
\cite{Xing04}. Note that the scale of $M^{}_{\rm R}$ can naturally
be much higher than the electroweak scale $v$, because those
right-handed Majorana neutrinos are $\rm SU(2)^{}_L$ singlets and
their corresponding mass term is not subject to the magnitude of
$v$.

Without loss of generality, it is convenient to choose a flavor
basis in which $M^{}_l$ is diagonal, real and positive (i.e., the
flavor and mass eigenstates of three charged leptons are
identified with each other). Then we concentrate on the
$(3+n)\times (3+n)$ neutrino mass matrix in Eq. (4), where
$M^{}_{\rm D}$ is a $3\times n$ matrix and $M^{}_{\rm R}$ is an
$n\times n$ matrix. The typical number of $n$ is of course $n=3$,
but $n=2$ is also a very interesting option as discussed in the
so-called minimal seesaw models \cite{FGY}. One may diagonalize
the symmetric $(3+n)\times (3+n)$ neutrino mass matrix by use of a
unitary transformation matrix:
\begin{equation}
\left ( \matrix{ V  &  R \cr S  &  U \cr} \right )^\dagger \left (
\matrix{ {\bf 0}     & M^{}_{\rm D} \cr M^T_{\rm D}    & M^{}_{\rm
R} \cr} \right ) \left ( \matrix{ V  &  R \cr S  &  U \cr} \right
)^* = \left ( \matrix{ \overline{M}^{}_\nu  &  {\bf 0} \cr {\bf 0}
& \overline{M}^{}_{\rm R} \cr} \right ) \; ,
\end{equation}
where $R$, $S$, $U$ and $V$ are the $3\times n$, $n\times 3$,
$n\times n$ and $3\times 3$ sub-matrices, respectively;
$\overline{M}^{}_\nu$ and $\overline{M}^{}_{\rm R}$ denote the
diagonal $3\times 3$ and $n\times n$ mass matrices with
eigenvalues $m^{}_i$ and $M^{}_j$ (for $i=1,2,3$ and $j= 1,
\cdots, n$), respectively. Eq. (5) yields
\begin{equation}
S^\dagger M^T_{\rm D} R^* + V^\dagger M_{\rm D} U^* + S^\dagger
M_{\rm R} U^* \; =\; {\bf 0} \; , ~~~~
\end{equation}
and
\begin{eqnarray}
\overline{M}^{}_\nu & = & S^\dagger M^T_{\rm D}V^* + V^\dagger
M_{\rm D} S^* + S^\dagger M_{\rm R} S^* \; ,
\nonumber \\
\overline{M}_{\rm R} & = & U^\dagger M^T_{\rm D} R^* + R^\dagger
M_{\rm D} U^* + U^\dagger M_{\rm R} U^* \; .
\end{eqnarray}
With the help of Eq. (6), $S^\dagger$ can be expressed as
\begin{equation}
S^\dagger = -V^\dagger M^{}_{\rm D}M^{-1}_{\rm R} \left [ {\bf 1}
+ M^T_{\rm D}R^*(U^*)^{-1}M^{-1}_{\rm R} \right ]^{-1} \; .
\end{equation}
Combining Eqs. (7) and (8), we arrive at
\begin{eqnarray}
V \overline{M}^{}_\nu V^T & = & - M^{}_{\rm D}M^{-1}_{\rm
R}M^T_{\rm D} + \Delta^{}_V  \; ,
\nonumber \\
U \overline{M}^{}_{\rm R} U^T & = & M^{}_{\rm R} + \Delta^{}_U \; ,
\end{eqnarray}
where
\begin{eqnarray}
\Delta^{}_V & = & M^{}_{\rm D} M^{-1}_{\rm R} M^T_{\rm D} R^* R^T -
M^{}_{\rm D}M^{-1}_{\rm R} (U^\dagger)^{-1} R^\dagger M^{}_{\rm D}
S^* V^T \; ,  \nonumber \\
\Delta_U & = & M^T_{\rm D}R^* U^T - M^{}_{\rm R}S^* S^T \; .
\end{eqnarray}
It is worth remarking that we have made no approximation in
obtaining Eqs. (9) and (10). Because the $(3+n)\times (3+n)$
transformation matrix in Eq. (5) is unitary, its four sub-matrices
satisfy the following conditions:
\begin{eqnarray}
V^\dagger V + S^\dagger S & = & VV^\dagger + RR^\dagger = {\bf 1}
\; , \nonumber \\
U^\dagger U + R^\dagger R & = & UU^\dagger + SS^\dagger = {\bf 1} \; ;
\end{eqnarray}
and
\begin{eqnarray}
V^\dagger R + S^\dagger U & = & VS^\dagger + RU^\dagger = {\bf 0}
\; ,
\nonumber \\
R^\dagger V + U^\dagger S & = & SV^\dagger + UR^\dagger = {\bf 0}
\; .
\end{eqnarray}
Obviously, $U$, $V$, $R$ and $S$ are in general not unitary.

Note that $V$ is just the MNS neutrino mixing matrix. To see this
point more clearly, one may re-express ${\cal L}^{}_{\rm cc}$ in
Eq. (1) by using the mass eigenstates of three charged leptons and
those of $(3+n)$ neutrinos. The latter can be denoted as
$\nu^{}_i$ (for $i=1,2,3$) and $N^{}_n$ (for $i=1, \cdots, n$),
which are related to $(\nu^{}_e, \nu^{}_\mu, \nu^{}_\tau)$ through
\begin{equation}
\left ( \matrix{ \nu^{}_e \cr \nu^{}_\mu \cr \nu^{}_\tau \cr}
\right )^{}_{\rm L} =  V \left ( \matrix{ \nu^{}_1 \cr \nu^{}_2
\cr \nu^{}_3 \cr} \right )_{\rm L} + R \left ( \matrix{ N^{}_1 \cr
\vdots \cr N^{}_n \cr} \right )_{\rm L} \; .
\end{equation}
Then ${\cal L}^{}_{\rm cc}$ reads
\begin{equation}
-{\cal L}^{}_{\rm cc} \; =\; \frac{g}{\sqrt{2}} \left [
\overline{(e, ~\mu, ~\tau)^{~}_{\rm L}} ~ V \gamma^\mu \left
(\matrix{ \nu^{}_1 \cr \nu^{}_2 \cr \nu^{}_3 \cr} \right )_{\rm L}
W^-_\mu ~ + ~ \overline{(e, ~\mu, ~\tau)^{~}_{\rm L}} ~ R
\gamma^\mu \left (\matrix{ N^{}_1 \cr \vdots \cr N^{}_n \cr}
\right )_{\rm L} W^-_\mu \right ] + {\rm h.c.} \; .
\end{equation}
We observe that $V$ enters the charged-current interactions
between three charged leptons $(e, \mu, \tau)$ and three
well-known light neutrinos $(\nu^{}_1, \nu^{}_2, \nu^{}_3)$, while
$R$ is relevant to the charged-current interactions between $(e,
\mu, \tau)$ and $(N^{}_1, \cdots, N^{}_n)$. Thus $V$ is the MNS
matrix. The unitarity of $V$ is naturally violated, due to the
presence of non-vanishing $R$ and $S$. A preliminary upper bound
on the matrix elements of $R$ is at the ${\cal O}(10^{-3})$ level,
extracted from some precise electroweak data \cite{B}. In the
limit of $R\rightarrow {\bf 0}$ and $S \rightarrow {\bf 0}$, $V$
turns out to be exactly unitary.

\vspace{0.5cm}

\framebox{\large\bf 3} ~ For simplicity, we denote the mass scales
of $M^{}_{\rm R}$ (or $\overline{M}^{}_{\rm R}$) and $M^{}_{\rm
D}$ as $M^{}_0$ and $m^{}_0$, respectively. Of course, $M^{}_0 \gg
v$ and $m^{}_0 \lesssim v$ are naturally expected in almost all
the reasonable extensions of the SM. The smallness of
$m^{}_0/M^{}_0$ implies that the sub-matrices $R$ and $S$ are
strongly suppressed in magnitude. This point can straightforwardly
be observed from Eq. (8), which approximates to
\begin{equation}
S^\dagger \; \approx \; -V M^{}_{\rm D} M^{-1}_{\rm R} \; \sim \;
{\cal O}(m^{}_0/M^{}_0) \; .
\end{equation}
On the other hand, Eq. (5) yields
\begin{equation}
R \; = \; +M^{}_{\rm D} U^* \overline{M}^{-1}_{\rm R} \; \sim \;
{\cal O}(m^{}_0/M^{}_0) \; .
\end{equation}
These results, together with Eqs. (11) and (12), lead to
\begin{eqnarray}
V^\dagger V & \approx & V V^\dagger \approx {\bf 1} \; ,
\nonumber \\
U^\dagger U & \approx & U U^\dagger \approx {\bf 1} \; ,
\end{eqnarray}
which hold up to ${\cal O}(m^2_0/M^2_0)$. Then we arrive at the
light Majorana neutrino mass matrix
\begin{equation}
M^{}_\nu \; \equiv \; V \overline{M}^{}_\nu V^T \; \approx \; -
M^{}_{\rm D} M^{-1}_{\rm R}M^T_{\rm D} \;
\end{equation}
and the heavy Majorana neutrino mass matrix $M^{}_{\rm R} \approx
U \overline{M}^{}_{\rm R} U^T$ from Eq. (9) as two good
approximations. Eq. (18) is just the well-known (Type-I) seesaw
relation between $M^{}_\nu$ and $M^{}_{\rm R}$ \cite{SS}. It
indicates that the mass scale of three light neutrinos is of
${\cal O}(m^2_0/M^{}_0)$. In other words, the smallness of three
left-handed neutrino masses is essentially attributed to the
largeness of $n$ right-handed neutrino masses.

To illustrate how the unitarity of $V$ or $U$ is slightly violated
in a more explicit way, let us consider the simplest seesaw model
with only a single heavy right-handed Majorana neutrino (i.e.,
$n=1$). In this special case, $M^{}_{\rm R} = M^{}_0$ holds
\footnote{Because the rank of $M^{}_{\rm R}$ equals one, the
seesaw relation in Eq. (18) implies that $M^{}_\nu$ is also a
rank-one neutrino mass matrix. Thus two of its three mass
eigenvalues must vanish, leading to a vanishing neutrino
mass-squared difference. This result is certainly in contradiction
with current solar and atmospheric neutrino oscillation
experiments. In other words, the canonical seesaw model with a
single heavy right-handed Majorana neutrino is not realistically
viable.}.
The $3\times 1$ matrix $R$ and the $1\times 3$ matrix $S$ can be
written as
\begin{equation}
R \; =\; \left ( \matrix{ r^{}_x \cr r^{}_y \cr r^{}_z \cr} \right
) \; , ~~~~~~ S^T \; =\; \left ( \matrix{ s^{}_x \cr s^{}_y \cr
s^{}_z \cr} \right ) \; .
\end{equation}
Then we obtain
\begin{eqnarray}
R^\dagger R & = & |r^{}_x|^2 + |r^{}_y|^2 + |r^{}_z|^2 \; \equiv
\; |r|^2 \; ,
\nonumber \\
S S^\dagger & = & |s^{}_x|^2 + |s^{}_y|^2 + |s^{}_z|^2 \; \equiv
\; |s|^2 \; .
\end{eqnarray}
Note that $|r| \sim |s| \sim m^{}_0/M^{}_0$ holds. In view of Eq.
(11), the departure of $U^\dagger U$ or $U U^\dagger$ from unity
is at the ${\cal O}(m^2_0/M^2_0)$ level. On the other hand,
\begin{eqnarray}
R R^\dagger & = & \left ( \matrix{ |r^{}_x|^2 & r^{}_x r^*_y &
r^{}_x r^*_z \cr r^*_x r^{}_y & |r^{}_y|^2 & r^{}_y r^*_z \cr
r^*_x r^{}_z & r^*_y r^{}_z & |r^{}_z|^2 \cr} \right ) \; ,
\nonumber \\
S^\dagger S & = & \left ( \matrix{ |s^{}_x|^2 & s^*_x s^{}_y &
s^*_x s^{}_z \cr s^{}_x s^*_y & |s^{}_y|^2 & s^*_y s^{}_z \cr
s^{}_x s^*_z & s^{}_y s^*_z & |s^{}_z|^2 \cr} \right ) \; .
\end{eqnarray}
It becomes obvious that the magnitude of each matrix element of $R
R^\dagger$ or $S^\dagger S$ is at most of ${\cal O}(|r|^2)$ or
${\cal O}(|s|^2)$. Hence the deviation of $V^\dagger V$ or $V
V^\dagger$ from the $3\times 3$ identity matrix is also at the
${\cal O}(m^2_0/M^2_0)$ level.

Given $m^{}_0 \sim 100$ GeV and $m^2_0/M^{}_0 \sim 0.1$ eV, one
may easily obtain $M^{}_0 \sim 10^{14}$ GeV. The latter is just
the typical mass scale of heavy right-handed Majorana neutrinos in
most of the realistic seesaw models. This estimate implies that
the magnitude of $R$ or $S$ is of ${\cal O}(m^{}_0/M^{}_0) \sim
{\cal O}(10^{-12})$. Hence the above-obtained seesaw formula is
valid up to a high accuracy of ${\cal O}(m^2_0/M^2_0) \sim {\cal
O}(10^{-24})$. Noticeably, the unitarity of the $3\times 3$ MNS
matrix is only violated at the ${\cal O}(10^{-24})$ level in such
a canonical seesaw scenario. It is therefore very safe to neglect
the extremely tiny ${\cal O}(m^2_0/M^2_0)$ correction to both
$M^{}_\nu$ and $V$.

The accuracy of Eq. (18) should be highlighted, because this
seesaw formula was naively regarded as an approximation of ${\cal
O}(m^{}_0/M^{}_0)$. Our instructive analysis shows that its
validity is actually up to ${\cal O}(m^2_0/M^2_0)$. Furthermore,
the unitarity violation of $V$ or $U$ can only take place at the
${\cal O}(m^2_0/M^2_0)$ level. That is why the $3\times 3$ MNS
neutrino mixing matrix is almost unitary in the realistic seesaw
models.

\vspace{0.5cm}

\framebox{\large\bf 4} ~ Note that Eq. (18) is usually referred to
as the Type-I seesaw relation. A somehow similar relation, the
so-called Type-II seesaw formula, can be derived from the
generalized lepton mass term
\begin{equation}
-{\cal L}^{\prime\prime}_{\rm lepton} \; =\; \overline{e^{}_{\rm
L}} M^{}_l e^{}_{\rm R} + \frac{1}{2} \overline{(\nu^{}_{\rm L},
~N^{\rm c}_{\rm R} )} \left ( \matrix{ M^{}_{\rm L}     &
M^{}_{\rm D} \cr M^T_{\rm D}    & M^{}_{\rm R} \cr} \right ) \left
( \matrix{ \nu^{\rm c}_{\rm L} \cr N^{}_{\rm R} \cr} \right ) +
{\rm h.c.} \; ,
\end{equation}
where $M^{}_{\rm L}$ may result from a new Yukawa interaction term
which violates the $\rm SU(2)^{}_L \times U(1)^{}_Y$ gauge
symmetry \cite{Review}. The mass scale of $M^{}_{\rm L}$ is likely
to be much lower than the electroweak scale $v$. Following the
strategies outlined above, one may diagonalize the $(3+n) \times
(3+n)$ neutrino mass matrix in Eq. (22) and arrive at the
effective light Majorana neutrino mass matrix
\begin{equation}
M^{}_\nu \; \equiv \; V \overline{M}^{}_\nu V^T \; \approx \;
M^{}_{\rm L} - M^{}_{\rm D} M^{-1}_{\rm R} M^T_{\rm D} \; ,
\end{equation}
where $V$ is the $3\times 3$ MNS neutrino mixing matrix. This
result is just the Type-II seesaw relation. Since the mass scale
of $M^{}_{\rm L}$ is expected to be smaller than that of
$M^{}_{\rm D}$ in those realistic models \cite{Review}, Eq. (23)
is valid up to the accuracy of ${\cal O}(m^2_0/M^2_0)$. The
unitarity of $V$ is also violated at the ${\cal O}(m^2_0/M^2_0)$
level, analogous to the Type-I seesaw case.

We conclude that the $3\times 3$ MNS matrix $V$, which appears in
the leptonic charged-current weak interactions, must not be
exactly unitary in the canonical (Type-I) and Type-II seesaw
models. Its unitarity violation is extremely small, as required by
the models themselves to reproduce the correct mass scale of light
Majorana neutrinos. Nevertheless, the unitarity of $V$ could be
more significantly violated by other sources of new physics (e.g.,
the existence of additional heavy charged leptons or light sterile
neutrinos \cite{Guo}). We remark that testing the unitarity of
$V$, both its normalization conditions and its orthogonality
relations \cite{Zhang}, is one of the important experimental tasks
to be fulfilled in the future neutrino factories and super-beam
facilities.

\acknowledgments{We would like to thank W.L. Guo and J.W. Mei for
useful discussions. This work is supported in part by the National
Natural Science Foundation of China.}


\begin{thebibliography}{99}
 \bibitem{SNO} SNO Collaboration, Q.R. Ahmad {\it et al.},
Phys. Rev. Lett. {\bf 89}, 011301 (2002).

\bibitem{SK} For a review, see: C.K. Jung {\it et al.},
Ann. Rev. Nucl. Part. Sci. {\bf 51}, 451 (2001).

\bibitem{KM} KamLAND Collaboration, K. Eguchi {\it et al.},
Phys. Rev. Lett. {\bf 90}, 021802 (2003); CHOOZ Collaboration, M.
Apollonio {\it et al.}, Phys. Lett. B {\bf 420}, 397 (1998); Palo
Verde Collaboration, F. Boehm {\it et al.}, Phys. Rev. Lett. {\bf
84}, 3764 (2000).

\bibitem{K2K} K2K Collaboration, M.H. Ahn {\it et al.},
Phys. Rev. Lett. {\bf 90}, 041801 (2003).

\bibitem{Review} For recent reviews with extensive references, see, e.g.,
H. Fritzsch and Z.Z. Xing, Prog. Part. Nucl. Phys. {\bf 45}, 1
(2000); S.F. King, Rept. Prog. Phys. {\bf 67}, 107 (2004); G.
Altarelli and F. Feruglio, New J. Phys. {\bf 6}, 106 (2004); R.N.
Mohapatra {\it et al.}, hep-ph/0510213.

\bibitem{MNS} Z. Maki, M. Nakagawa, and S. Sakata,
Prog. Theor. Phys. {\bf 28}, 870 (1962).

\bibitem{CKM} N. Cabibbo, Phys. Rev. Lett. {\bf 10}, 531 (1963);
M. Kobayashi and T. Maskawa, Prog. Theor. Phys. {\bf 49}, 652
(1973).

\bibitem{Xing04} See, e.g., Z.Z. Xing, Int. J. Mod. Phys. A {\bf 19}, 1
(2004).

\bibitem{SS} P. Minkowski, Phys. Lett. B {\bf 67}, 421 (1977);
T. Yanagida, in {\it Proceedings of the Workshop on Unified Theory
and the Baryon Number of the Universe}, edited by O. Sawada and A.
Sugamoto (KEK, Tsukuba, 1979), p. 95; M. Gell-Mann, P. Ramond, and
R. Slansky, in {\it Supergravity}, edited by F. van Nieuwenhuizen
and D. Freedman (North Holland, Amsterdam, 1979), p. 315; S.L.
Glashow, in {\it Quarks and Leptons}, edited by M.
L$\rm\acute{e}vy$ {\it et al.} (Plenum, New York, 1980), p. 707;
R.N. Mohapatra and G. Senjanovic, Phys. Rev. Lett. {\bf 44}, 912
(1980).

\bibitem{FGY} P. Frampton, S.L. Glashow, and T. Yanagida, Phys.
Lett. B {\bf 548}, 119 (2002). See, also, W.L. Guo and Z.Z. Xing,
Phys. Lett. B {\bf 583}, 163 (2004); J.W. Mei and Z.Z. Xing, Phys.
Rev. D {\bf 69}, 073003 (2004).

\bibitem{B} E. Nardi, E. Roulet, and D. Tommasini, Phys. Lett. B
{\bf 344}, 225 (1995).

\bibitem{Guo} See, e.g., W.L. Guo and Z.Z. Xing, Phys. Rev. D {\bf
65}, 073020 (2002); Phys. Rev. D {\bf 66}, 097302 (2002).

\bibitem{Zhang} See, e.g., H. Zhang and Z.Z. Xing, Eur. Phys. J. C
{\bf 41}, 143 (2005); Z.Z. Xing and H. Zhang, Phys. Lett. B {\bf
618}, 131 (2005).
\end{thebibliography}
\end{document}